\documentclass[a4paper,11pt]{article}
\usepackage{pos}
\usepackage{subfigure}

\newcommand{\be}{\begin{eqnarray}}
\newcommand{\ee}{\end{eqnarray}}
\newcommand{\nee}{\nonumber\end{eqnarray}}

\newcommand{\drbar}{{\overline{\rm DR}}}

\newcommand{\msg}    {m_{\ti g}}

\def\be            {\begin{equation}}
\def\ee            {\end{equation}}
\def\bea            {\begin{eqnarray}}
\def\eea            {\end{eqnarray}}

\def\l               {\lambda}

\def\x               {\chi}
\def\ti              {\tilde}

\def\st              {\ti t}
\def\sc              {\ti c}
\def\sbot            {\ti b}
\def\ch              {\ti \x^\pm}

\def\nt              {\ti \x^0}
\def\sg              {\ti g}

\def\su                  {\ti{u}}

\def\sd                  {\ti{d}}
\def\ss                  {\ti{s}}

\definecolor{darkgreen}{rgb}{0,.5,0}

\title{The $h(125)$ decays to $c \bar c$, $b \bar b$, $b \bar s$, $\gamma \gamma$ and $g g$ in the light of the MSSM with quark flavor violation}
\ShortTitle{The $h(125)$ decays in the MSSM with quark flavor violation}

\author*[a]{Keisho Hidaka}
\author[b]{Helmut Eberl}
\author[b]{Elena Ginina}

\affiliation[a]{Department of Physics, Tokyo Gakugei University,\\
  Koganei, Tokyo 184-8501, Japan}

\affiliation[b]{Institut f\"ur Hochenergiephysik der \"Osterreichischen Akademie
der Wissenschaften,\\
A-1050 Vienna, Austria}

\emailAdd{hidaka@u-gakugei.ac.jp}
\emailAdd{helmut.eberl@oeaw.ac.at}
\emailAdd{elena.ginina@oeaw.ac.at}

\abstract{We study the Higgs boson decays $h^0 \to c \bar{c}, b \bar{b}, b \bar{s}, \gamma \gamma, 
g g$ in the Minimal Supersymmetric Standard Model (MSSM) with general quark flavor violation (QFV), 
identifying the $h^0$ as the Higgs boson with a mass of 125 GeV. We compute the widths of the 
$h^0$ decays to $c \bar c, b \bar b, b \bar s (s \bar b)$ at full one-loop level. 
For the $h^0$ decays to photon photon and gluon gluon we compute the widths at NLO QCD level. 
We perform a systematic MSSM parameter scan respecting all the relevant constraints, i.e. 
theoretical constraints from vacuum stability conditions and experimental constraints, 
such as those from K- and B-meson data and electroweak precision data, as 
well as recent limits on Supersymmetric (SUSY) particle masses and the 
125 GeV Higgs boson data from LHC experiments. From the parameter scan, we find 
that the deviations of these MSSM widths from the Standard Model (SM) values can be 
quite sizable. All of these large deviations in the $h^0$ decays are due to large 
scharm-stop mixing, large scharm/stop involved trilinear couplings 
$T_{U23}, T_{U32}, T_{U33}$, large sstrange-sbottom mixing, and large 
sstrange/sbottom involved trilinear couplings $T_{D23}, T_{D32}, T_{D33}$. 
International Linear Collider (ILC) can observe such large deviations from 
the SM at high signal significance. In case the deviation pattern shown here 
is really observed at ILC, then it would strongly suggest the discovery of 
QFV SUSY (the MSSM with QFV).}

\FullConference{%
  *** The European Physical Society Conference on High Energy Physics (EPS-HEP2021), ***\\
  *** 26-30 July 2021 ***\\
  *** Online conference, jointly organized by Universität Hamburg and the research center DESY ***
}

\begin{document}

\begin{flushright}
HEPHY-PUB 1029/21
\end{flushright}

\maketitle

\section{Introduction}
What is the SM-like Higgs boson discovered at LHC? 
It can be the SM Higgs boson. 
It can be a Higgs boson of New Physics. 
This is one of the most important issues in the present particle physics field. 
Here we study a possibility that it is the lightest Higgs boson $h^0$ of the 
Minimal Supersymmetric Standard Model (MSSM), focusing on the decays $h^0(125) 
\to c \bar c, b \bar b , b \bar s, \gamma \gamma, g g$. This work is based on the 
update of our previous papers \cite{h02cc,h02bb,h02gagagg} and contains 
significant (substantial) new findings. 

\section{Key parameters of the MSSM}
Key parameters in this study are the quark flavor violating (QFV) parameters 
$M^2_{Q_{u}23} (\simeq M^2_{Q23})$, $M^2_{U23}$, $T_{U23}$, $T_{U32}$, 
$M^2_{Q23}$, $M^2_{D23}$, $T_{D23}$ and $T_{D32}$ which describe 
the $\ti{c}_L - \ti{t}_L$, $\ti{c}_R - \ti{t}_R$, $\ti{c}_R - \ti{t}_L$, 
$\ti{c}_L - \ti{t}_R$, $\ti{s}_L - \ti{b}_L$, $\ti{s}_R - \ti{b}_R$, 
$\ti{s}_R - \ti{b}_L$, and $\ti{s}_L - \ti{b}_R$ mixing, respectively. 
The quark flavor conserving (QFC) parameters $T_{U33}$ and $T_{D33}$ 
which induce the $\ti{t}_L - \ti{t}_R$ and $\ti{b}_L - \ti{b}_R$ mixing, 
respectively, also play an important role in this study. 
All the parameters in this study are assumed to be real, except the 
CKM matrix. We also assume that R-parity is conserved and that the 
lightest neutralino $\nt_1$ is the lightest SUSY particle (LSP). 

\section{Constraints on the MSSM}
\label{Constraints}
In our study we perform a MSSM-parameter scan respecting all the relevant 
constraints, i.e. the theoretical constraints from vacuum stability 
conditions and the experimental constraints, such as those 
from $K$- and $B$-meson data and electroweak precision data, 
as well as recent limits on SUSY particle masses and 
the $H^0$ mass and coupling data from LHC experiments. 
Here $H^0$ is the discovered SM-like Higgs boson which we 
identify as the lightest $CP$ even neutral Higgs boson $h^0$ in the MSSM.
The details of these constraints are summarized in Ref. \cite{C7paper}.

\section{Parameter scan for $h^0(125) \to c \bar c, b \bar b , b \bar s$}
\label{ParameterScan}
We perform the MSSM parameter scan for the decay widths $\Gamma(h^0 
\to c \bar c)$, $\Gamma(h^0 \to b \bar b)$ and $\Gamma(h^0 \to b \bar s)$ 
computed at full 1-loop level in the MSSM with QFV respecting all the 
relevant constraints mentioned above. 
Concerning squark generation mixings, we only consider the mixing between 
the second and third generation of squarks.
We generate the input parameter points by using random numbers 
in the ranges shown in Table~\ref{table1}, where some parameters are fixed 
as given in the last box. All input parameters are $\drbar$ parameters 
defined at scale Q = 1 TeV, except $m_A$(pole) which is the pole mass of 
the $CP$ odd Higgs boson $A^0$. The parameters that are not shown 
explicitly are taken to be zero. We don't assume a GUT relation for the 
gaugino masses $M_1$, $M_2$, $M_3$.

\begin{table}[h!]
\footnotesize{
\caption{
Scanned ranges and fixed values of the MSSM parameters (in units of GeV or GeV$^2$, 
except for $\tan\beta$). The parameters that are not shown explicitly are 
taken to be zero. $M_{1,2,3}$ are the U(1), SU(2), SU(3) gaugino mass parameters, respectively.}
\begin{center}
\begin{tabular}{|c|c|c|c|c|c|}
    \hline
\vspace*{-0.3cm}
& & & & &\\
\vspace*{-0.3cm}
     $\tan\beta$ & $M_1$ &  $M_2$ & $M_3$ & $\mu$ &  $m_A(pole)$\\ 
& & & & &\\
    \hline
\vspace*{-0.3cm}
& & & & &\\
\vspace*{-0.3cm}
     10 $\div$ 80 & $100 \div 2500$ & $100 \div 2500$  & $2500 \div 5000$ & $100 \div 2500$ & $1350 \div 6000$\\
& & & & &\\
    \hline
    \hline
\vspace*{-0.3cm}
& & & & &\\
\vspace*{-0.3cm}
      $ M^2_{Q 22}$ & $ M^2_{Q 33}$ &  $|M^2_{Q 23}| $ & $ M^2_{U 22} $ & $ M^2_{U 33} $ &  $|M^2_{U 23}| $\\ 
& & & & &\\
     \hline
\vspace*{-0.3cm}
& & & & &\\
\vspace*{-0.3cm}
      $2500^2 \div 4000^2$ & $2500^2 \div 4000^2$ & $< 1000^2$  & $1000^2 \div 4000^2$ & $600^2 \div 3000^2$& $ < 2000^2$\\
& & & & &\\
    \hline
    \hline
\vspace*{-0.3cm}    
& & & & &\\
\vspace*{-0.3cm}      
      $ M^2_{D 22} $ & $ M^2_{D 33}$ &  $ |M^2_{D 23}|$ & $|T_{U 23}|  $ & $|T_{U 32}|  $ &  $|T_{U 33}|$\\ 
& & & & &\\
    \hline
\vspace*{-0.3cm}      
& & & & &\\
\vspace*{-0.3cm}  
       $ 2500^2 \div 4000^2$ & $1000^2 \div 3000^2 $ & $ < 2000^2$  & $< 4000 $ & $ < 4000$& $< 5000 $\\
& & & & &\\
 \hline 
\multicolumn{6}{c}{}\\[-3.6mm]  
\cline{1-4}
\vspace*{-0.3cm}      
     & & & \\
\vspace*{-0.3cm}      
     $ |T_{D 23}| $ & $|T_{D 32}|  $ &  $|T_{D 33}|$ &$|T_{E 33}| $\\ 
     & & & \\
    \cline{1-4}
\vspace*{-0.3cm}      
     & & & \\
\vspace*{-0.3cm}      
     $< 3000 $ & $< 3000 $& $ < 4000$& $ < 500$\\
     & & & \\
    \cline{1-4}
\end{tabular}\\[3mm]
\begin{tabular}{|c|c|c|c|c|c|c|c|c|}
    \hline
\vspace*{-0.3cm}      
    & & & & & & & &\\
\vspace*{-0.3cm}      
    $M^2_{Q 11}$ & $M^2_{U 11} $ &  $M^2_{D 11} $ & $M^2_{L 11}$ & $M^2_{L 22} $ &  $M^2_{L 33}$ & $M^2_{E 11}$&$M^2_{E 22}$ & $M^2_{E 33} $\\ 
    & & & & & & & &\\
    \hline
\vspace*{-0.3cm}      
    & & & & & & & &\\
\vspace*{-0.3cm}      
    $4500^2$ & $4500^2$ & $4500^2$  & $1500^2$ & $1500^2$ & $1500^2$& $1500^2$& $1500^2$&$1500^2$\\
    & & & & & & & &\\
    \hline
\end{tabular}
\end{center}
\label{table1}
}
\end{table}

\indent
From 377180 input points generated in the scan 3208 points survived 
all the constraints. We show these survival points in all scatter plots 
in this article.

\section{$h^0(125) \to c \bar c, b \bar b , b \bar s$ in the MSSM}
\begin{figure*}[t!]
\centering
  \subfigure[]{
  {\mbox{\resizebox{3.8cm}{!}{\includegraphics{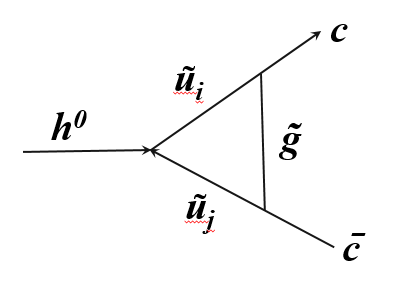}}}}
  \label{h02cc_gluino_loop}}
  \subfigure[]{
  {\mbox{\resizebox{4cm}{!}{\includegraphics{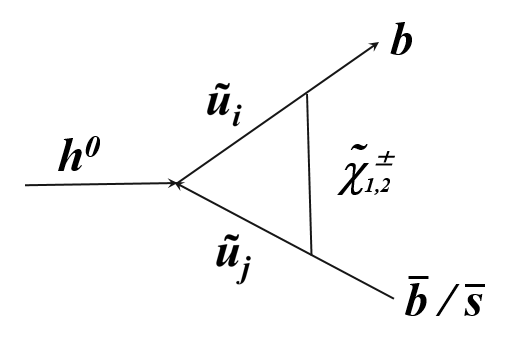}}}}
  \label{h02bb_chargino_loop}}
  \subfigure[]{
  {\mbox{\resizebox{4cm}{!}{\includegraphics{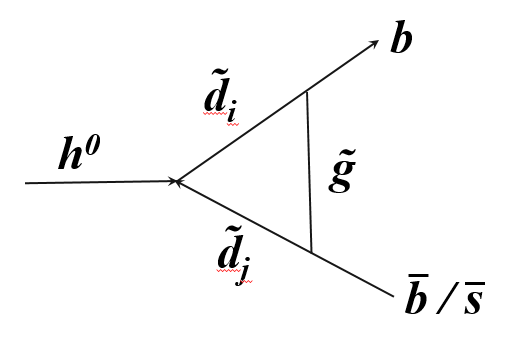}}}}
  \label{h02bb_gluino_loop}}
\caption{
(a) The $\su_i$-$\sg$ loop corrections to $\Gamma(h^0 \to c \bar c)$, 
(b) the $\su_i$-$\ch_{1,2}$ loop and (c) the $\sd_i$-$\sg$ loop corrections 
to $\Gamma(h^0 \to b \, \, \bar b / \bar s)$.
}
\label{1-loop_diag_to_h0_decay}
\end{figure*}

We compute the decay widths $\Gamma(h^0 \to c \bar c)$, 
$\Gamma(h^0 \to b \bar b)$ and $\Gamma(h^0 \to b \bar s)$ at full 
1-loop level in the $\overline{DR}$ renormalization scheme in the 
MSSM with QFV \cite{h02cc,h02bb}. 
We find that large squark trilinear couplings $T_{U23,32,33}$, 
$T_{D23,32,33}$, large $M^2_{Q23}$, $M^2_{U23}$, $M^2_{D23}$, 
large bottom Yukawa coupling $Y_b$ for large $\tan\beta$, and 
large top Yukawa coupling $Y_t$ can lead to large MSSM 1-loop 
corrections to the widths of these decays. 
This is mainly due to the following reasons:
The lighter up-type squarks $\su_{1,2,3}$ are strong $\sc_{L,R}$ - 
$\st_{L,R}$ mixtures for large $M^2_{Q23}$, $M^2_{U23}$, 
$T_{U23,32,33}$. The lighter down-type squarks $\sd_{1,2,3}$ 
are strong $\ss_{L,R}$ - $\sbot_{L,R}$ mixtures for large 
$M^2_{Q23}$, $M^2_{D23}$, $T_{D23,32,33}$.
Here note that $|T_{U23,32,33}|$, the sizes of which are controlled 
by $Y_t$ due to the vacuum stability conditions, can be large 
because of large $Y_t$. Similarly $|T_{D23,32,33}|$, the sizes of 
which are controlled by $Y_b$ due to the vacuum stability 
conditions, can be large thanks to large $Y_b$ for large 
$\tan\beta$ \cite{C7paper}. In the following we assume 
these setups.\\
Main MSSM 1-loop corrections to $\Gamma(h^0 \to c \bar c)$ stem 
from the up-type squarks ($\su_{1,2,3}$) - gluino ($\sg$) loops 
at the decay vertex which have $h^0-\su_i-\su_j$ couplings 
containing $H^0_2-\sc_R-\st_L$, $H^0_2-\sc_L-\st_R$, 
$H^0_2-\st_L-\st_R$ couplings, i.e., $T_{U23,32,33}$ 
(see Fig. \ref{h02cc_gluino_loop}). Hence, 
large trilinear couplings $T_{U23,32,33}$ can enhance the 
$h^0-\su_i-\su_j$ couplings, which results in enhancement of 
the MSSM 1-loop corrections to $\Gamma(h^0 \to c \bar c)$ due 
to the $\su_i$-$\sg$ loops, leading to large deviation of the 
MSSM width $\Gamma(h^0 \to c \bar c)$ from its SM value.
Main MSSM 1-loop corrections to $\Gamma(h^0 \to b \bar b)$ and 
$\Gamma(h^0 \to b \bar{s}/\bar{b} s)$ stem from 
(i) $\su_{1,2,3}$ - chargino ($\ch_{1,2}$) loops at the decay vertex 
which have $h^0-\su_i-\su_j$ couplings to be enhanced by  
large $T_{U23,32,33}$ (see Fig. \ref{h02bb_chargino_loop}) and 
(ii) $\sd_{1,2,3}$ - $\sg$ loops at the decay vertex which have 
$h^0-\sd_i-\sd_j$ couplings containing $H^0_1-\ss_R-\sbot_L$, 
$H^0_1-\ss_L-\sbot_R$, $H^0_1-\sbot_L-\sbot_R$ couplings, i.e., 
$T_{D23,32,33}$ (see Fig. \ref{h02bb_gluino_loop}). 
Hence large trilinear couplings $T_{U23,32,33}$ 
and $T_{D23,32,33}$ can enhance the MSSM 1-loop corrections 
to $\Gamma(h^0 \to b \bar b)$ and $\Gamma(h^0 \to b \bar{s}/\bar{b} s)$  
due to the $\su_i$ - $\ch_{1,2}$ and $\sd_i$ - $\sg$ loops, 
leading to large deviation of the MSSM widths $\Gamma(h^0 \to b \bar b)$ 
and $\Gamma(h^0 \to b \bar{s}/\bar{b} s)$ from their SM values.

We define the deviation of the MSSM width from the SM width as:
\be
DEV(X) \equiv \frac{\Gamma(h^0 \to X \bar X)_{MSSM}}{\Gamma(h^0 \to X \bar X)_{SM}} - 1 \, \, (X=c,b)
  \label{DEVX}
\ee

\noindent DEV(X) is related with the coupling modifier 
$\kappa_X \equiv C(h^0 X \bar X)_{MSSM}/C(h^0 X \bar X)_{SM}$ as $DEV(X)=\kappa_X^2 -1$. 
We compute the decay widths $\Gamma(h^0 \to X \bar X)$ (X=c,b) at full 
1-loop level in the $\overline{DR}$ renormalization scheme in the MSSM 
with QFV using Fortran codes developed by us \cite{h02cc,h02bb}.  

\begin{figure*}[t!]
\centering
 {\mbox{\resizebox{7.0cm}{!}{\includegraphics{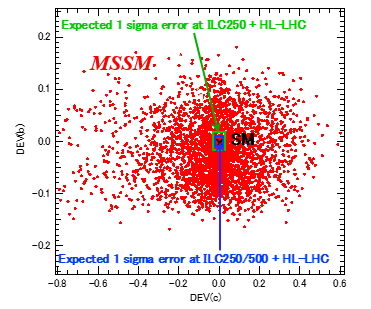}}}}
\caption{
The scatter plot in the DEV(c)-DEV(b) plane obtained 
from the MSSM parameter scan described in Section \ref{ParameterScan}. 
"X" marks the SM point. 
The green and blue box indicate the expected 1$\sigma$ error at 
[ILC250 + HL-LHC] and [ILC250/500 + HL-LHC], respectively. 
}
\label{DEVc2DEVb}
\end{figure*}

\begin{figure*}[t!]
\centering
 {\mbox{\resizebox{7.0cm}{!}{\includegraphics{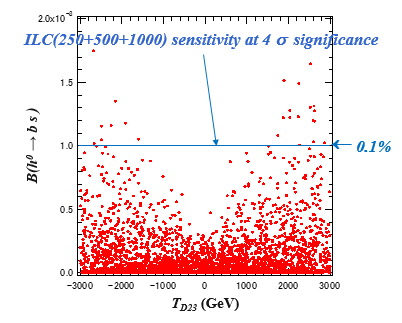}}}}
\caption{
The scatter plot in the $T_{D23}$-$B(h^0 \to b s)$ plane obtained 
from the MSSM parameter scan described in Section \ref{ParameterScan}. 
The blue horizontal line indicates the ILC(250+500+1000) sensitivity 
of $\sim$0.1\% at 4$\sigma$ signal significance.
}
\label{BRbs}
\end{figure*}

In Fig. \ref{DEVc2DEVb} we show the scatter plot in the DEV(c)-DEV(b) plane 
obtained from the MSSM parameter scan described above (see Table \ref{table1}), 
respecting all the relevant constraints shown in Section \ref{Constraints}.
From Fig. \ref{DEVc2DEVb} we see that DEV(c) and DEV(b) can be quite 
large simultaneously: DEV(c) can be as large as $\sim\pm 50 \%$ and 
DEV(b) can be as large as $\sim\pm 20 \%$.
ILC together with HL-LHC can observe such large deviations from SM at 
high significance \cite{ILC_Higgs}:
The expected 1$\sigma$ error of DEV(c) is $\Delta$DEV(c) =(3.60\%, 2.40\%, 1.58\%) 
and that of DEV(b) is $\Delta$DEV(b) =(1.98\%, 1.16\%, 0.94\%) 
at (ILC250, ILC250/500, ILC250/500/1000) together with HL-LHC, respectively 
(see Fig. \ref{DEVc2DEVb}).\\
\indent As for the explicitly QFV decay branching ratio 
$B(h^0 \to b s) \equiv B(h^0 \to b \bar{s}) + B(h^0 \to \bar{b} s)$, 
from our MSSM parameter scan we find that it can be as large as $\sim 0.17\%$ 
in the MSSM with QFV (see also \cite{Heinemeyer}) while it is almost zero in the SM.
On the other hand, the ILC250/500/1000 sensitivity to this branching 
ratio could be $\sim 0.1\%$ at 4$\sigma$ signal significance \cite{Tian}. 
Note that LHC and HL-LHC sensitivity should not be so good due to 
huge QCD background.
In Fig. \ref{BRbs} we show the scatter plot in the $T_{D23}$-$B(h^0 \to b s)$ 
plane obtained from the MSSM parameter scan described above 
(see Table \ref{table1}), respecting all the relevant constraints shown 
in Section \ref{Constraints}. 
From Fig. \ref{BRbs} we see that there is a strong correlation between 
$T_{D23}$ and $B(h^0 \to b s)$: $B(h^0 \to b s)$ can be large for large 
$|T_{D23}|$ being the size of the $\ti{s}_R - \ti{b}_L$ mixing parameter. 
We have obtained a similar result for the scatter plot in the 
$T_{D32}$-$B(h^0 \to b s)$ plane with $T_{D32}$ being the 
$\ti{s}_L - \ti{b}_R$ mixing parameter. 

%

\section{$h^0(125) \to \gamma \gamma, g g$ in the MSSM}
For the $h^0$ decays to $\gamma \gamma$ and $g g$ we compute the widths 
at NLO QCD level \cite{h02gagagg}. We perform the MSSM parameter scan respecting all 
the relevant theoretical and experimental constraints as mentioned above.
From the parameter scan, we find the followings \cite{h02gagagg}: \\ 
(i) DEV($\gamma$) and DEV($g$) can be sizable simultaneously: 
    DEV($\gamma$) and DEV($g$) can be as large as $\sim +4\%$ 
    and $\sim -15\%$, respectively.\\
(ii) There is a very strong correlation between DEV($\gamma$) 
    and DEV($g$), which is due to the fact that the "stop"-loop 
    (i.e. stop-scharm mixture loop) contributions dominate 
    the two DEVs.\\
(iii) The deviation of the width ratio 
    $\Gamma(h^0 \to \gamma \gamma)/\Gamma(h^0 \to g g)$ 
    in the MSSM from the SM value can be as large as $\sim +20\%$.\\
(iv) ILC250/500 together with HL-LHC can observe such large 
    deviations from SM at high significance \cite{ILC_Higgs}.\\
The scatter plots in the DEV($\gamma$)-DEV($g$) plane obtained 
from the MSSM parameter scan are shown in Ref. \cite{h02gagagg}.

\section{Conclusion}
We have studied the decays $h^0(125)\to c \bar c, b \bar b, 
b \bar s, \gamma \gamma, g g$ in the MSSM with QFV. 
Performing a systematic MSSM parameter scan respecting all 
of the relevant theoretical and experimental constraints, 
we have found the followings:\\
(A) DEV(c) and DEV(b) can be very large simultaneously:
    DEV(c) and DEV(b) can be as large as $\sim\pm$50\% 
    and $\sim\pm$20\%, respectively.\\
(B) The deviation of the width ratio 
   $\Gamma(h^0 \to b \bar b)/\Gamma(h^0 \to c \bar c)$ 
   in the MSSM from the SM value can exceed +100\%.\\
(C) $B(h^0 \to b s)$ can be as large as $\sim$0.17\% in the MSSM 
   while ILC250/500/1000 sensitivity could be $\sim$0.1\% 
   at 4$\sigma$ signal significance.\\
(D) DEV($\gamma$) and DEV($g$) can be sizable simultaneously: 
    DEV($\gamma$) and DEV($g$) can be as large as $\sim +4\%$ 
    and $\sim -15\%$, respectively.\\
(E) The deviation of the width ratio 
    $\Gamma(h^0 \to \gamma \gamma)/\Gamma(h^0 \to g g)$ 
    in the MSSM from the SM value can be as large as $\sim +20\%$.\\
(F) There is a very strong correlation between DEV($\gamma$) 
    and DEV($g$).\\
(G) All of these large deviations in the $h^0$(125) decays 
    are due to (i) large $\ti{c} - \ti{t}$ mixing and large $\ti{c}/\ti{t}$ 
    involved trilinear couplings $T_{U23}, T_{U32}, T_{U33}$, 
    (ii) large $\ti{s} - \ti{b}$ mixing and large $\ti{s}/\ti{b}$ 
     involved trilinear couplings $T_{D23}, T_{D32}, T_{D33}$, 
    (iii) large bottom Yukawa coupling $Y_b$ for large $\tan\beta$, 
     and large top Yukawa coupling $Y_t$. \\
\indent ILC together with HL-LHC can observe such large deviations from SM 
at high significance. In case the deviation pattern shown here is 
really observed at ILC, then it would strongly suggest the discovery 
of QFV SUSY (the MSSM with QFV).

\end{document}